\documentclass[prl,twocolumn]{revtex4}

\usepackage{amsmath}
\usepackage{amssymb}
\usepackage{graphicx}

\begin{document}
\title{Woven Nematic Defects, Skyrmions and the Abelian Sandpile Model}
\author{Thomas Machon}
\affiliation{H. H. Wills Physics Laboratory, University of Bristol, Bristol BS8 1TL, UK}
\author{Gareth P. Alexander}
\affiliation{Department of Physics and Centre for Complexity Science, University of Warwick, Coventry, CV4 7AL, UK}
\begin{abstract}
We show that a fixed set of woven defect lines in a nematic liquid crystal supports a set of non-singular topological states which can be mapped on to recurrent stable configurations in the Abelian sandpile model or chip-firing game. The physical correspondence between local Skyrmion flux and sandpile height is made between the two models. Using a toy model of the elastic energy, we examine the structure of energy minima as a function of topological class and show that the system admits domain wall Skyrmion solitons. 

\end{abstract}
\maketitle

The interplay of periodicity and topology can lead to particularly rich phenomonology, as evidenced by both electronic~\cite{hasan10} and mechanical systems~\cite{mao18}. In the liquid crystal literature, blue phases with their networks of defect lines provide a classic example~\cite{wright89}. Many other periodic topological structures, typically stabilised through surface geometry and topology have been found in liquid crystalline systems, with strong analogies to chiral magnetic systems~\cite{machon16a}. Following early work on effective interactions between colloids in liquid crystals~\cite{lubensky98, poulin97} regular arrays of defects and colloids were found~\cite{musevic06}, with similar techniques being used to create large systems of entangled and knotted defect lines~\cite{tkalec11}.  Defect arrays have also been created by immersing sheets with regular arrays of punctures into a liquid crystal host~\cite{tran16}, or with arrays of toron excitations~\cite{ackerman17} as well as in numerical studies in cholesterics~\cite{fukuda11b}.

\begin{figure}
\begin{center}
\includegraphics{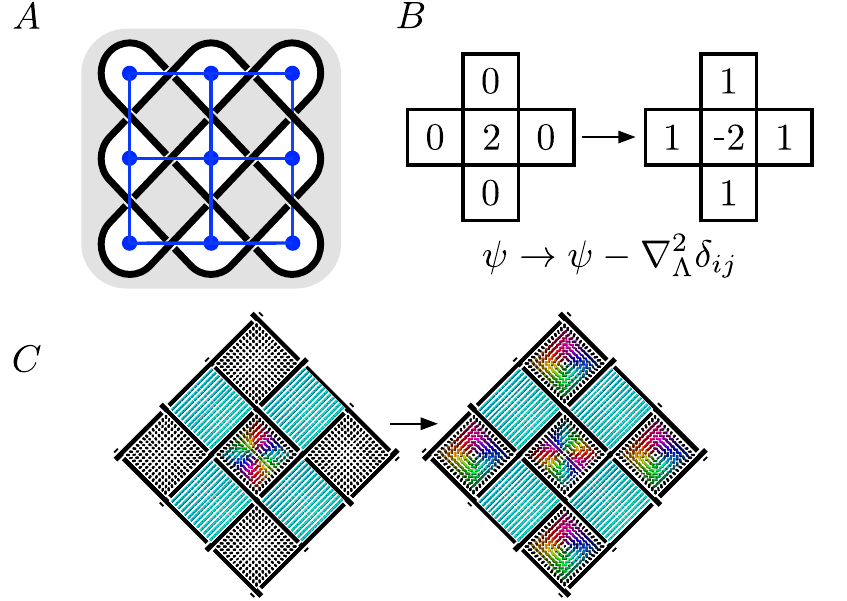}
\end{center}
\caption{$A$: $3 \times 3$ example of the defect arrays we consider. The defect lattice $\mathcal{L}$ is shown in black. The surface on which the $\{\alpha \}$ cycles reside is shown in grey and the lattice $\Lambda$ is shown in blue. More generally, one should think of the equivalent $n \times n$ array with alternating over and under crossings. $B$: Illustration of an elementary hopping move. The Skyrmion field $\psi$ may be deformed by any field $\nabla^2_\Lambda g$, here $g$ chosen to be a $\delta$-function at site $(i,j)$. $C$: Sketch depicting a possible realisation of the hopping move in $B$.}
\label{fig:array}
\end{figure}

In this paper we study topological physics in a periodic system of a different kind: Skyrmions in nematic liquid crystals entangled with a fixed lattice of woven defect lines (Fig.~\ref{fig:array}). We note that systems containing arrays of Skyrmions have been studied numerically in cholesterics~\cite{fukuda11} and experimentally in Skyrmion bags~\cite{foster18}. We show that the topologically allowed transitions between different Skyrmion configurations in a fixed lattice of defects can be expressed in terms of a simple set of allowed hopping moves for Skyrmion charge. By considering these allowed moves as an equivalence relation, we show there is a duality between topological classes of Skyrmions on such a lattice and stable recurrent states of the Abelian sandpile model on a graph associated to the defect lattice. Physically, one identifies sandpile height with local Skyrmion charge in the lattice and the allowed Skyrmion hopping moves correspond directly to generalised collapses in the Abelian sandpile model. Mathematically this result can be thought of in the context of correspondences between the recurrent states of the Abelian sandpile model and graph Laplacians~\cite{dhar90}, between graph Laplacians and rooted spanning trees (via the matrix-tree theorem), between rooted spanning trees and branched double covers~\cite{kauffman}, and between branched double covers and nematic defect topology~\cite{machon14,machon16}. Motivated by the analogy with sandpile height, we construct a toy model for the energetics of these states. We investigate the structure of groundstates in each topological class for a $3\times 3$ square lattice. We then consider the toy model in the case of large systems which exhibits soliton structures consisting of lines of Skyrmions. In a physical realisation of the system we have in mind, the defect lines could be inclusions, fibres or colloids, which simulate disclination lines in the surrounding nematic texture in a similar manner to the imprinting of defect lines in the blue phases using polymers~\cite{kikuchi02}, or in the creation of colloids that either mimic disclination lines~\cite{cavallarro13} or have desired topological properties~\cite{senyuk13}.

Skyrmions appear as topological distortions in nematic liquid crystals and have an integer charge, $q$. The Skyrmion flux through a surface $\Sigma$ may be computed via an integral of the topological charge density in terms of the unit magnitude director field ${\bf n}$ as
\begin{equation}
q = \frac{1}{8 \pi} \int_\Sigma dA_i \epsilon_{ijk} \epsilon_{abc}n_a \partial_j n_b \partial_k n_c,
\label{eq:skch}
\end{equation}
which is an integer if $\Sigma$ is closed. There are two important observations that one can make from \eqref{eq:skch}. The first is that measuring Skyrmion charge is done via a surface integral, so that Skyrmions naturally live on surfaces. Therefore, if the Skyrmion charge is localised to a region on a given surface then in three dimensions the Skyrmion charge becomes confined to a tube-like object. The second observation is that under the action of the nematic inversion symmetry $\mathbf{n} \to - \mathbf{n}$, $q \to - q$, so that locally Skyrmion charge is only defined up to sign. Mathematically, this represents the action of the fundamental group $\pi_1(\mathbb{R} \mathbb{P}^2)$ on $\pi_2(\mathbb{R} \mathbb{P}^2)$, where $\mathbb{R} \mathbb{P}^2 = S^2/(x \sim -x)$ is the nematic groundstate manifold~\cite{mermin79,alexander12}.

Consider a Skyrmion in a fixed defect array ${\cal L}$ such as the square woven lattice of Fig.~\ref{fig:array}. For a Skyrmion in a free system, there is a translational symmetry so the Skyrmion may move around~\cite{ackerman17a}. In the lattice of defects there are not only geometric but topological obstructions for the Skyrmion to move around the lattice. As we show below, Skyrmions may only move around the lattice according to generalisations of the hopping rule shown in Fig.~\ref{fig:array}. These hopping rules originate in the interaction between Skyrmions and line defects~\cite{alexander12} and its relation to the $q \to -q$ and ${\bf n} \to - {\bf n}$ symmetry. Around a nematic disclination line the director field is non-orientable -- as can be seen in the typical $\pm 1/2$ profiles. Tracking ${\bf n}$ along a circuit around a disclination line one finds that it realises the transformation ${\bf n} \to -{\bf n}$. This changes the charge of the Skyrmion. More exotic topological aspects of nematics can be understood in a similar manner~\cite{machon14,machon16}. A way of visualising this is as follows: If a pair of point defects are nucleated in a nematic, then a Skyrmion distortion is created that connects them~\cite{machon16a,chen2013}. Consider now dragging one of the point defects along a path that entangles several line defects and then meets back up and annihilates with the original defect, leaving behind a Skyrmion entangled with the disclination lattice. Up to smooth deformations of the texture there are a finite number of topologically inequivalent ways that this may occur, indexed by a twisted cohomology group~\cite{machon16} and from the perspective of algebraic topology the distinct topological states of Skyrmions entangled with a fixed set of disclination lines are given by elements in the set
\begin{equation}
H^2(\mathbb{R}^3 \setminus \mathcal{L} ; \; \mathbb{Z}^\omega) / (x \sim - x) ,
\label{eq:cohomology}
\end{equation}
where $H^2(\mathbb{R}^3 \setminus \mathcal{L} ; \; \mathbb{Z}^\omega)$ is the twisted cohomology group with coefficient system $\mathbb{Z}^\omega$ given by the integers along with the map $\gamma: \pi_1(\mathbb{R}^3 \setminus \mathcal{L}) \to \mathbb{Z}_2$ which sends each meridian of $\mathcal{L}$ to $-1 \in \mathbb{Z}_2$. If the disclination line is a knot or link, then the size of this cohomology group is a knot invariant, the knot determinant, which is equal to the Alexander polynomial evaluated at $-1$~\cite{machon14,machon16}. 

\begin{figure}
\begin{center}
\includegraphics{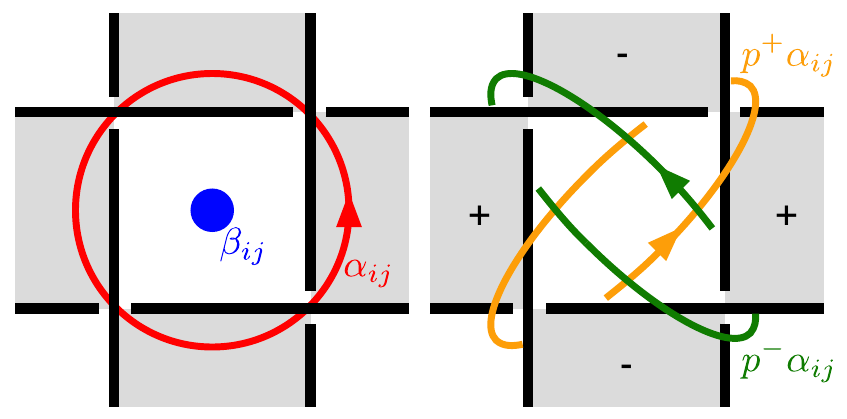}
\end{center}
\caption{Calculation of $H^2(\mathbb{R}^3 \setminus \mathcal{L} ; \; \mathbb{Z}^\omega)$ for the regular array of defect lines. The push-offs of each cycle $\alpha_{ij}$ from the gray surface in both the positive and negative directions are computed in terms of the cycles $\beta_{ij}$. As shown, the lattice vectors make an angle of $45^\circ$ with the horizontal/vertical axes.}
\label{fig:calc}
\end{figure}

Our goal is to study the possible ways of entangling Skyrmions with a fixed lattice of woven defect lines such as the one shown in Fig.~\ref{fig:array}, given by \eqref{eq:cohomology}. Mathematically this amounts to studying topological equivalence classes of textures, {\it i.e.} free homotopy classes of maps~\cite{mermin79} $[\Omega, \mathbb{RP}^2]$ where $\Omega = \mathbb{R}^3 \setminus \mathcal{L}$, where $\mathcal{L}$ is the defect set. Importantly this means that all the results presented here assume that no other defects interact with the system, our classification holds only as long as no additional defects are created and the topology of the defect set does not change. As discussed above, the defect lines in the array may be thought of as generated by inclusions, and we assume that the boundary conditions are such that the director field ${\bf n}$ is free on the surface of the defect lines (up to topological class). 

To compute $\eqref{eq:cohomology}$ for a set of line defects $\mathcal{L}$ such as the ones shown in Fig.~\ref{fig:array} one first draws a checkerboard surface $\Sigma$ for $\mathcal{L}$, as shown in Fig.~\ref{fig:calc}. One then writes a basis $\{ \alpha \}$ of cycles on $\Sigma$ and a dual basis of cycles in $\mathbb{R}^3 \setminus \Sigma$, which are bases for the homology groups  $H_1(\Sigma; \mathbb{Z})$ and $H_1(\mathbb{R}^3 \setminus \Sigma ; \mathbb{Z})$ respectively. These bases satisfy the condition
\begin{equation}
\textrm{Lk}(\alpha_i, \beta_j) = \delta_{ij},
\end{equation}
as illustrated in Fig.~\ref{fig:calc}, where the vertical $\beta$ cycles meet up at infinity. From a physical perspective, $\beta_{ij}$ represents Skyrmion flux through the site $(i,j)$, the $\alpha$ cycles encode the topological information about how this Skymrion flux can be moved around the defect lattice. As drawn in Fig.~\ref{fig:calc}, $\Sigma$ is orientable, and one can consider the push-offs, $p^\pm \alpha_{ij}$ of cycles on $\Sigma$ in either the positive or negative direction, which can be expressed as linear combinations of cycles in $\{\beta \}$. In the case of Fig.~\ref{fig:calc} it is readily shown that 
\begin{align}
p^+ \alpha_{ij} &= -\beta_{i+1,j} - \beta_{i-1,j} + 2 \beta_{ij}, \\ 
p^- \alpha_{ij} &= -\beta_{i,j+1} - \beta_{i,j-1} + 2 \beta_{ij}. 
\end{align}
The group $H^2(\mathbb{R}^3 \setminus \mathcal{L} ; \; \mathbb{Z}^\omega)$ is then given by integer combinations of the $\beta_{ij}$ along with the equivalence relations given by $p^+ \alpha_{ij} = - p^- \alpha_{ij}$ for each $(i,j)$. Considered as a linear map we have $p^+ +  p^-  = -\nabla_\Lambda^2$, the graph Laplacian of a black graph, $\Lambda$, of $\mathcal{L}$ (Fig.~\ref{fig:array} A). This defines the equivalence relation on the possible sets of Skyrmion fluxes, and consequently the topologically distinct ways of entangling Skyrmions with this array are given by integer-valued functions $\psi: \Lambda \to \mathbb{Z}$, with two functions equivalent if they differ by any integer-valued function $f$ in the image of the graph Laplacian, $f=\nabla^2_\Lambda g$, so that
\begin{equation}
\psi \sim \psi + \nabla^2_\Lambda g.
\label{eq:oursand}
\end{equation}
Finally, accounting for the ${\bf n} \to - {\bf n}$ symmetry leads to an identification of $\psi$ with $-\psi$. The set of states thus has the structure of $\tilde G_\Lambda/ x \sim -x$, where $\tilde G_\Lambda$ is the quotient group $\mathbb{Z}^{|\Lambda|} / \nabla^2_\Lambda \mathbb{Z}^{|\Lambda |}$. For a finite lattice, we may write 
\begin{equation}
\tilde G_\Lambda  = \mathbb{Z} \oplus G_\Lambda,
\label{eq:full}
\end{equation} where $G_\Lambda$ is a finite Abelian group and $\mathbb{Z}$ measures the charge of the entire lattice taken as a single object. Although here we focus on the square woven lattice shown in Fig.~\ref{fig:array}, other lattices are possible and described by the same general framework. For example, one can take the Kagome lattice, considered as families of straight lines, and make the vertices alternating over- and under- crossings; $\Lambda$ is then the honeycomb lattice. 

The group $G_\Lambda$ is known as the sandpile group for the lattice $\Lambda$, its order is given by $\textrm{pdet}(\nabla^2_\Lambda) / |\Lambda|$. In statistical mechanics it indexes stable recurrent states in the Abelian sandpile model~\cite{bak87} or chip-firing games~\cite{bjorner91,biggs99}. The Abelian sandpile model (see Ref.~\cite{dhar99} for a review) on a square lattice $\Lambda$ associates to each lattice $i$ site a non-negative height $h_i$. Sites with $h_k \geq 4$ are termed unstable and collapse according to the rule
\begin{equation}
h_i \to h_i - (\nabla^2_\Lambda)_{ij} \delta_{jk},
\label{eq:sand}
\end{equation}
The dynamics of the model is specified by incrementing $h$ at a random site by one, and then collapsing sites until the system is stable. To ensure this process terminates on a finite lattice, one designates a node as the sink which may not collapse.

\begin{figure}
\begin{center}
\includegraphics{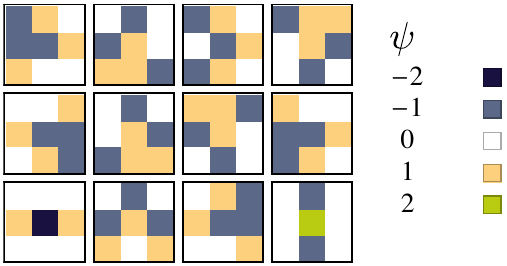}
\end{center}
\caption{The 12 lowest energy configurations for the state $\phi = (0,4)$, with energy $6$. This is the highest groundstate energy and degeneracy of any state on the $3 \times 3$ lattice.}
\label{fig:something}
\end{figure}

In a seminal paper~\cite{dhar90} Dhar showed that the set of recurrent stable configurations of this system are given by the sandpile group, $G_\Lambda$. The analogy between \eqref{eq:oursand} and \eqref{eq:sand} is clear. In the sandpile model, the function $h$ represents sandpile height at each lattice site. We are therefore motivated to consider $\psi$ as an analogous field, measuring local Skymrion charge. While such an equivalence is not strictly possible from a topological perspective, as $\psi$ is not a topological invariant (Skyrmions are extended objects and cannot be rigorously localised in general), it is likely that a physical realisation of this system would display sufficient structural regularity for $\psi$ to be defined in an {\it ad hoc} manner. There is some subtlety originating in the $x \sim -x$ symmetry of the nematic. This can be alleviated~\cite{machon16} by choosing a branch cut, realised as a spanning surface for the defect array (as shown by the grey surface in Fig.~\ref{fig:calc}), after which the director field may be oriented, and a signed field $\psi$ may be defined. Note that in our system, $\psi$ can take all integer values rather than non-negative as in the case of $h$.  \eqref{eq:oursand} then becomes an algebraic description of the allowed rules for moving Skyrmion charge around the lattice. In the sandpile model, the presence of the sink node ensures that the state space is finite, we can enforce a similar condition by demanding the total charge in the defect array is zero, so that $\sum_i \psi_i =0$. In a physical system this can be enforced by anchoring conditions at the boundary of the cell containing the defect array, for example. In this case, the number of states is finite, and the connection to the sandpile model further allows us to estimate the number of topological states for a square lattice of size $N$. In this case, one may show that the size of the group (and hence the number of topological states) scales as $|G_\Lambda | \sim e^{N^2 s}$, where
\begin{equation}
s = \frac{1}{(2\pi)^2}\int_{T^2} \log(2 - \cos \theta_1 - \cos \theta_2) d \theta_1 d \theta_2,
\label{eq:ent}
\end{equation}
is the entropy per site in the sandpile model~\cite{dhar90} as well as 2D phantom polymers at their critical point~\cite{kardar}. 

We now consider a system realising these topological states. We assume that the configuration can be labelled by a function $\psi: \Lambda \to \mathbb{Z}$ describing the local Skyrmion charge at each lattice site. We then wish to give an expression for the elastic energy of the configuration with a given $\psi$. This depends on the director field configuration ${\bf n}$. The nematic is symmetric under ${\bf n} \to - {\bf n}$, whereas the Skyrmion charge changes sign, so that any expression for the energy of the system must be symmetric under $\psi \to - \psi$. If we assume that the director field is in a local elastic energy minimum at each lattice site, then to lowest order we might try to write down a coarse-grained energy for the system as a function of the topological charge:
\begin{equation}
F = \frac{K}{2}\sum_{i \in \Lambda} \psi_i^2,
\label{eq:toy_en}
\end{equation}
where $K$ is an energy scale. Given \eqref{eq:toy_en}, it is natural to ask for the groundstate configuration within each topological class. An explicit representative for each state may be obtained from the Smith normal form~\cite{dhar95}, $P^{-1} \nabla^2_\Lambda Q=M$, where $M$ is diagonal. Let $\tau_i$ denote the $i^\textrm{th}$ invariant factor of $\nabla^2_\Lambda$, then given a group element 
\begin{equation}
\phi = (\phi_1, \ldots, \phi_n) \in \bigoplus_{i=1}^n \mathbb{Z}_{\tau_i}
\end{equation}
a function $\psi : \Lambda \to \mathbb{Z}$ representing the class $\phi$ can be written as $\tilde P^T\phi$, where $\tilde P^T$ consists of the columns of $P^T$ corresponding to the non-trivial elementary divisors. In general we may therefore write configurations in the form
\begin{equation}
\psi = \tilde P^T\phi + \nabla_\Lambda^2 \chi,
\label{eq:vec}
\end{equation}
where $\chi : \Lambda \to \mathbb{Z}$ is an arbitrary integer-valued function. To minimise the energy in a given topological state, we choose $\chi$ in \eqref{eq:vec} that minimises \eqref{eq:toy_en} for a given $\phi$. 

If $\nabla_\Lambda^2$ is Laplacian integral then the groundstates for each topological class may be found easily. The $3 \times 3$ lattice (as shown in Fig.~\ref{fig:array} $A$) is the largest Laplacian integral square lattice and so will serve as our example. In this case $G_\Lambda = \mathbb{Z}_8 \oplus \mathbb{Z}_{24}$, so without the nematic symmetry there are $192$ topological states. Accounting for the nematic and lattice symmetries reduces this to 42. \eqref{eq:vec} gives $\phi=(a,b)$, with $a \in \mathbb{Z}_8$ and $b \in \mathbb{Z}_{24}$. In general, there is groundstate degeneracy for a given $\phi$. Fig.~\ref{fig:something} shows the 12 lowest energy configurations for the state $\phi = (0,4)$, with energy $6$, the highest groundstate energy and degeneracy of any state on the $3 \times 3$ lattice. Note there are configurations in this class with $|\psi| \leq 1$, and this is true for all classes on the $3 \times 3$ lattice. Skyrmions are typically observed with charge $\pm 1$ and one may ask whether it is possible to use the topology of the defect lattice to create higher order charges. Mathematically, this is equivalent to asking whether
\begin{equation}
m_\Lambda(\phi) = \inf_\chi \| \tilde P^T \phi + \nabla^2_\Lambda \chi \|_{\infty}
\label{eq:svp}
\end{equation}
is equal to 1. For an arbitrary knotted or linked defect array this is not true~\cite{machon14,machon16}, as evidenced by the $(4,4)$ torus link. More generally computing $m_\Lambda(\phi)$ is related to the shortest lattice vector problem in the $\infty$-norm which is known to be NP-hard~\cite{vanemdeboas81}. On probabilistic grounds it likely that $m_\Lambda(\phi)=1$, the number of configurations with $|\psi| \leq n$ and $\sum \psi =0$ is equal to the $|\Lambda|^{\text{th}}$ central $(2n+1)^\text{st}$ multinomial coefficient. In particular, for $n=1$, the central trinomial coefficients, $T_{|\Lambda |}$, may be written in terms of Legendre polynomials as
\begin{equation}
T_{|\Lambda |} = (-3)^{|\Lambda|/2}P_{|\Lambda |}\big( (-3)^{-1/2} \big) \sim \frac{3^{|\Lambda|+1/2}}{2 \sqrt{2 |\Lambda|}} .
\end{equation}
As $\ln 3>s$ \eqref{eq:ent}, there are far more configurations with $|\psi|<1$ than there are topological classes.

\begin{figure}
\begin{center}
\includegraphics{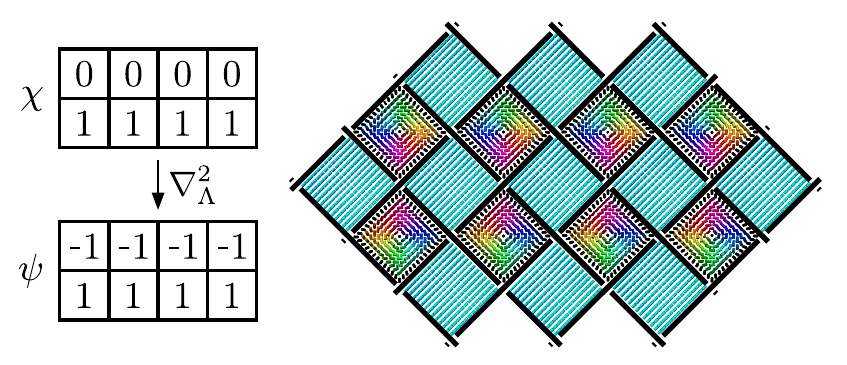}
\end{center}
\caption{Skyrmion domain wall. In an infinite system, $\chi$ has a set of constant groundstates, $\chi= c \in \mathbb{Z}$. At the interface between two such states where $c$ differs by one, the physical field $\psi$ contains two adjacent lines  of $+1$ and $-1$ Skyrmions forming a domain wall between the two groundstates.}
\label{fig:wall}
\end{figure}

Moving beyond these small-scale systems, one can instead consider a large system of the type shown in Fig.~\ref{fig:array}; a similar experimental system can be seen in Ref.~\cite{tran16}. On a general $N \times N$ lattice any state may be deformed, via moves of the form \eqref{eq:oursand}, so that it has support only on the boundary of the lattice. This can be done iteratively by setting $\psi$ to zero on successive Moore boundaries of an $n \times n$ sublattice, beginning with $n=1$. As such, in the infinite limit, the structure of the sandpile group becomes irrelevant, and the properties of the system are determined fully by the local move rules~\cite{note_gf}. We may thus write $\psi = \nabla^2_\Lambda \chi$ in this case so the energy becomes
\begin{equation}
F = \frac{K}{2} \langle \chi | \nabla_\Lambda^4 | \chi \rangle,
\label{eq:final_energy}
\end{equation}
where $\nabla^4_\Lambda$ is the biharmonic operator on $\Lambda$ and $\chi$ is an arbitrary integer valued function related to the physically observable field $\phi$ by $\phi = \nabla_\Lambda^2 \chi$. The groundstates of \eqref{eq:final_energy} are given by the kernel of $\nabla_\Lambda^2$ which in particular includes the constant functions, $\chi = c$. The system therefore admits domain wall Skyrmion solitons at the interface between $\chi = c_1$ and $\chi = c_2$, as indicated in Fig.~\ref{fig:wall}. Indeed, simple Monte Carlo simulations of the equilibrium statistical mechanics model with \eqref{eq:final_energy} as a Hamiltonian suggest the existence of a KT type transition involving the proliferation of such domain wall solitons. These domain walls consist of lines of $\pm 1$ Skyrmions, stabilised by the fixed defect array. 

\begin{acknowledgements}
It is a pleasure to acknowledge useful conversations with B.G. Chen. T. Machon would like to acknowledge funding from the NSF through grant DMR-1262047.
\end{acknowledgements}


\begin{thebibliography}{99}
\bibitem{note_gf} Computing the Green's function in the infinite case is an amusing exercise, the integer-valued requirement means that one must break the $D_4$ symmetry of the lattice. In particular, one may choose the Green's function such that it has support on a cone in $\mathbb{Z}^2$, with its values determined by the integer part of the resistance between appropriate points on an infinite lattice of unit magnitude resistors.
\end{thebibliography}
\end{document}